\documentclass[aps,amssymb,amsmath,
twocolumn,showpacs,superscriptaddress,
groupedaddress]{revtex4-1}

\usepackage{graphicx}
\usepackage{natbib}
\setcitestyle{square,numbers}
\usepackage{hyperref}
\usepackage{dcolumn}
\usepackage{bm}
\usepackage[table,xcdraw]{xcolor}
\usepackage{nth}
\usepackage[ruled,vlined]{algorithm2e}
\usepackage{framed}
\usepackage{listings}
\usepackage{tikz}
\usepackage{nicefrac, xfrac} 
\usetikzlibrary{shapes.geometric, arrows}
\usepackage[flushleft]{threeparttable}
\hypersetup{
  colorlinks,
  citecolor=red,
  linkcolor=blue,
  urlcolor=blue}

\begin{document}
\title{Solution of the Dirac$-$Coulomb equation using the Rayleigh$-$Ritz method.\\
Results for He$-$like atoms.}
\author{A. Ba{\u g}c{\i}}
\email{abagci@pau.edu.tr}
\affiliation{Department of Physics, Faculty of Sciences, Pamukkale University,\\{\c C}amlaraltı, K{\i}n{\i}kl{\i} Campus, 20160 Pamukkale, Denizli, Turkey}
\author{P. E. Hoggan}
\affiliation{Institute Pascal, UMR 6602 CNRS, Clermont-Auvergne University, 24 avenue des Landais BP 80026, 63178 Aubi\`ere Cedex, France}

\begin{abstract}
The Dirac$-$Coulomb equation for helium$-$like ions is solved using the iterative self$-$consistent field method, with Slater$-$type spinor orbitals as the basis. These orbitals inherently satisfy the kinetic$-$balance condition due to their coupling for both large$-$ and small$-$components. The $1/r_{12}$ Coulomb interaction is treated without constraints. Computations are carried out for total energies of atoms with nuclear charges up to $Z \leq 80$ using both minimal and extended basis sets. Variationally optimal values for orbital parameters are determined through the Rayleigh$-$Ritz variational principle. No manifestations related to the Brown$-$Ravenhall disease are found.
\begin{description}
\item[Keywords]
Dirac$-$Coulomb equation, Slater$-$type spinor orbitals, He$-$like atoms
\item[PACS numbers]
... .
\end{description}
\end{abstract}
\maketitle

\section{Introduction} \label{introduction}
The Dirac equation solution provides four$-$component spinors and a spectrum that with no finite lower bound \cite{Dirac_1930, Karwowski_2017_1}. This spectrum comprises three distinct intervals: 

i) An energy range $\left(-\infty, -m_{0}c^2 \right)$ that corresponds to the negative energy continuum states (the negative energy spectrum), 

ii) The interval $\left(-m_{0}c^2, m_{0}c^2\right)$ that contains the discrete spectrum of bound states, and 

iii) A range $\left(m_{0}c^2, \infty\right)$ that corresponds to positive continuum states (positive energy spectrum). 

The electrons and their anti$-$particles, positrons are characterized by the four$-$component spinor wave function. It can be written as:
\begin{align}\label{eq:1}
\Psi=
\begin{pmatrix}
\psi^{L1}\\
\psi^{L2}\\
\psi^{S1}\\
\psi^{S2}
\end{pmatrix}
\end{align}
The notation used in Eq. (\ref{eq:1}) is preferred for the positive energy solutions, corresponding to electronic states. The lower components ($S$) go to zero in the non$-$relativistic limit and the upper components ($L$) thus become a solution of the corresponding non$-$relativistic Schr{\"o}dinger equation. The labels $L, S$ are used to denote the terms \textit{large} and \textit{small}, respectively.

Previous work by the authors \citep{ABagci_2016, ABagci_2020} for one$-$electron systems shows that using Slater$-$type spinor orbitals (STSOs) in the algebraic solution of the Dirac$-$Coulomb differential equation via the linear combination of atomic spinors method (LCAS) \cite{Swirles_1935, Kim_1967} successfully eliminates the variational collapse \citep{Schwarz_1982_1, Schwarz_1982_2}. This assertion holds even when extended basis$-$set approximations are employed, where the quantum number $\kappa$ can have both positive and negative values. The STSOs are defined as \cite{ABagci_2020},
\begin{align}\label{eq:2}
	X_{nljm}\left(\zeta, \vec{r} \right)
	=\begin{pmatrix}
		\chi_{nljm}^{\beta 1}\left(\zeta, \vec{r}\right)
		\vspace{1.5 mm}\\
		\chi_{nljm}^{\beta -1}\left(\zeta, \vec{r}\right)
		\vspace{1.5 mm}\\
		\chi_{nljm}^{-\beta 1}\left(\zeta, \vec{r}\right)
		\vspace{1.5 mm}\\
		\chi_{nljm}^{-\beta -1}\left(\zeta, \vec{r}\right)
	\end{pmatrix},
\end{align}
here:
\begin{align}\label{eq:3}
	\chi_{nljm}^{\beta \eta}\left(\zeta, \vec{r}\right)
	=f^{\beta}_{nlj}(\zeta,r) \Omega_{ljm}^{\beta \eta} \left(\theta, \vartheta\right),
\end{align}
\begin{align}\label{eq:4}
	f^{\beta}_{nlj}(\zeta,r)
	=\left\{{A_{nlj}^{\beta}r^{n}+\zeta B_{nlj}^{\beta}r^{n+1}}\right\}e^{-\zeta r},
\end{align}
with, $\beta$ represents large$-$ and small$-$components of STSOs, $\left\lbrace n,l,j,m \right\rbrace$ are the principal, angular, total angular and secondary total angular momentum quantum numbers with $n \in \mathbb{R}^{+}$, $0\leq l \leq \lfloor n \rfloor-1$, $j=l\mp 1/2$, $-j \leq m \leq j$ and $\lfloor n \rfloor$ stands for the integer part of $n$ respectively. $\zeta$ are orbital parameters. \\
The $\Omega_{ljm}^{\beta \eta}$ are the spin $\frac{1}{2}$ spinor spherical harmonics \cite{Davydov_1976},
\begin{align}\label{eq:5}
	\Omega_{ljm}^{\beta \eta}\left(\theta, \vartheta\right)
	=\text{\large $a$}_{l_{\beta}jm \left( \eta \right)}
	\text{\large $\eta$}_{m\left( \eta \right)}
	Y_{l_{\beta} {m\left( \eta \right)}}\left(\theta, \vartheta\right),
\end{align}
where values of $l_{\beta}$ are determined by $l_{\beta}=j+\frac{\beta}{2}$, $\eta$ stands for the spherical part of each component of STSOs and $\eta_{m\left(\eta \right)}=\mathtt{i}^{\vert m\left(\eta \right)\vert-m\left(\eta \right)}$. The quantities $\text{\large $a$}$ are the Clebsch$-$Gordan coefficients \cite{Varshalovich_1988}. 

$Y_{lm_{l}}$ are complex spherical harmonics $(Y^{*}_{lm_{l}}=Y_{l-m_{l}})$,
\begin{align}\label{eq:6}
	Y_{l\vert m_{l} \vert}(\theta, \vartheta)
	=\frac{1}{2\pi}\mathcal{P}_{l|m_{l}|}\left(cos\theta \right)e^{\mathtt{i}m_{l}\vartheta}.
\end{align}
Their sign differs from Condon$-$Shortley phase by a factor: $(-1)^{m_{l}}$ \cite{Condon_1959}. $\mathcal{P}_{lm_{l}}(x)$ are the associated Legendre functions, $m_{l} \equiv m\left(\varepsilon \right)$ being magnetic quantum numbers.\\
The large$-$ and small$-$components in radial parts of STSOs are coupled as \cite{ABagci_2016,ABagci_2020},
\begin{multline}\label{eq:7}
	\frac{\partial}{\partial r}f_{n\kappa}^{\beta}\left(\zeta, r\right)
	=-\beta\frac{\kappa}{r}f_{n\kappa}^{\beta}\left(\zeta, r\right) \\
	+\left(\frac{\beta N_{n\kappa}-n-\delta_{\vert \kappa\vert \kappa}}{r}+ \zeta \right)f_{n\kappa}^{-\beta}\left(\zeta, r\right).
\end{multline}
This coupling applies to any real values of $n$ and allows calculations for atoms beyond critical nuclear charge $Z_{c}$, $Z_{c} \simeq 137.04$ \cite{ABagci_2020}. This is because the value of $n$ in STSOs can for instance be such that $n= \vert \kappa \vert$ is independent of the speed of light.

Contemplating increasingly sophisticated systems (i.e. those comprising atoms with more then one$-$electron) poses a formidable challenge in the derivation of a Lorentz covariant Dirac equation \cite{Karwowski_2017_1, Kutzelnigg_2012, Karwowski_2017_2}. A Hamiltonian to be used in relativistic quantum chemistry that is analogous to the non$-$relativistic many$-$electron Hamiltonian was derived nevertheless by using the Dirac's kinetic energy operator for every electron and the Coulomb approximation for electron$-$electron interaction \cite{Grant_2007}. In atomic units $\left(a.u.\right)$ it is given as:
\begin{align}\label{eq:8}
\hat{H}_{DC}=\sum_{i}\hat{h}_{D}\left(i\right)
+\sum_{i>j}\frac{1}{r_{ij}}.
\end{align}
$\hat{h}_{D}\left(i\right)$ is the Dirac Hamiltonian of a single $ith$ electron in interaction with fixed point$-$like nucleus:
\begin{align}\label{eq:9}
\hat{h}_{D}=
c\left(\vec{\alpha}.\hat{\vec{p}}\right)+\left(\beta-1\right)c^{2}-\frac{Z}{r_{i}},	
\end{align}
where $\vec{\alpha}$ stands for Pauli matrices, $\beta$ is $2\times 2$ Dirac matrix constructed from $2\times 2$ unit matrices. $\hat{\vec{p}}$ is the momentum operator, $c$ is the speed of light and $Z$ is the nuclear charge. $1/r_{ij}$ is the electrostatic interaction between $ith$ and $jth$ electrons.\\
The so called \textit{Brown$-$Ravenhall disease} \cite{Brown_1951} is an artifact of constructing a differential equation using Eq. (\ref{eq:8}). The discrete energy of bound states and continuum states of its one$-$electron part overlap. The discrete energy of bound states are coupled by Coulomb interaction term to the continuum states and that causes the continuum states to spread over entire spectrum from $\left(-\infty, \infty \right)$ (Brown$-$Ravenhall continuum) \cite{Karwowski_2017_2}.\\
It is worthwhile considering the non$-$relativistic formalism at this stage. The Schr{\"o}dinger$-$Coulomb differential equation can be written in a matrix form and solved as a generalized eigenvalue problem. It is solved in iterative steps, using the linear combination of atomic orbitals (LCAO) \cite{Roothaan_1951} method. An initial set of expansion coefficients is generated initially by neglecting the two$-$electron part. These coefficients are used to construct the Fock matrix. Solving the Fock matrix generates improved sets of expansion coefficients. This procedure is repeated until a convergence criterion is satisfied.\\
Application to the Dirac$-$Coulomb differential equation operates in a similar fashion, but this time, additional conditions between the steps are taken into account, due to unbounded property of the Dirac spectrum.
\begin{itemize}
\item{The first condition is to ensure that the physical ground state of the solution for one$-$electron part has a rigorous variational property.}
\end{itemize}
This is achieved by using for instance the \textit{restricted kinetic$-$balance} \cite{Stanton_1984, Dyall_1990, Sun_2011} condition:
\begin{align}\label{eq:10}
\psi^{S} = \frac{1}{2mc}\left(\vec{\sigma} . \hat{\vec{p}}\right) \psi^{L}
\end{align}
The small$-$components of the spinor basis can be derived from its large$-$components using Eq. (\ref{eq:10}). Note that several other types for kinetic$-$balance condition have been proposed to ensure variational stability \cite{Sun_2011}. All of them are based on the principle that the large$-$components of the spinor basis should be subsets of small$-$components \cite{Feagri_2005, Karwowski_2006, Dyall_2012}. The selection of which condition to use is determined by the specific objective of computation. Solution of the two$-$center Dirac equation via 'minimax' variation methods \cite{Chuluunbaatar_2021} for example, the spinor basis to be used in calculations may directly be chosen depending on the principle. Contribution of negative energy continuum states on the other hand, can significantly improve the accuracy \cite{Grant_2010, Tatewaki_2011}. Those interested in obtaining numerical results that properly represent the whole Dirac spectrum should therefore use a direct coupling between large$-$ and small$-$components as given in Eq. (\ref{eq:7}).\\
Optimization of variational parameters is another important task at this point. Variational stability should be satisfied over the whole range of orbital parameters $\zeta$, $\zeta >0$ in an arbitrary basis set approximation. The variational approach for a given calculation is determined by the kinetic$-$balance condition. The minimax method \cite{Talman_1986, Kolakowska_1996} is a standard choice if the condition does not strictly satisfy the symmetry of the Dirac's equation. Numerical results that prove flexibility of basis set approximations and correctly represent the whole Dirac spectrum for one$-$electron systems are unknown to us.
\begin{itemize}
\item{The second condition serves to guarantee that the Coulomb interaction term avoids the possibility of un$-$physical results when building the Fock matrix.}
\end{itemize}
The strategy to ignore appearance of Brown$-$Ravenhall continuum and hence eliminate the \textit{continuum dissolution} is to use positive energy projection $\left(\Lambda_{+}\right)$ of the operator and derive the so called \textit{no$-$pair projected} Dirac$-$Coulomb Hamiltonian \cite{Sucher_1980, Sucher_1985}.
\begin{align}\label{eq:11}
\hat{\mathcal{H}}_{DC}
=\sum_{i} \Lambda_{+}\hat{h}_{D}\left(i\right) \Lambda_{+}
+ \sum_{i>j} \Lambda_{+} \frac{1}{r_{ij}} \Lambda_{+}
\end{align}
Such an extraction that focuses only on correct representation of positive energy continuum states and discrete energy of bound states of electrons, unfortunately does not suffice to fully address the methodology of calculation for the many$-$electron case \cite{Grant_2007, Kutzelnigg_2012, Liu_2014}. The results obtained from calculations depend on the value of nuclear charge $Z$. Any perturbation or interaction introduces a contribution from negative energy continuum states. The no$-$pair Dirac$-$Coulomb Hamiltonian cannot completely eliminate the negative energy continuum states \cite{Grant_2007, Liu_2014}. The positive energy projection method can be further improved by implementing it through complex scaling of the particle coordinates \cite{Seba_1988,Pestka_2006} (see also references therein).
\begin{align}\label{eq:12}
r_{i} = r_{i}e^{i\theta}
\end{align}
where, $\theta$ is a real parameter. Demonstrations performed by using explicitly correlated basis sets may be found in \cite{Pestka_2006, Bylicki_2007}. Complex scaling has effectively remedied specific limitations associated with no$-$pair approximation, resulting in substantial improvements in both its theoretical rigor and overall accuracy for small nuclear charge \cite{Jeszenszki_2022}. This is due to the ability to separate the discrete energy of bound states from the negative energy continuum states. One should accept that the 'no-pair approximation' is a technical one, considered merely to avoid the continuum dissolution. It is initially non$-$self$-$adjoint, and methods like complex scaling are often employed \textit{post hoc} to enhance its self$-$adjoint properties. Studies from the mathematical stand point have been conducted in recent years on the domain, self$-$adjointness and spectrum of Dirac operators for interacting particles \cite{Deckert_2019}.

In previous work \cite{ABagci_2020} one of the authors proved that regularization of the Coulomb potential enables extension of the domain for one$-$electron Dirac$-$Coulomb Hamiltonian, achieving self$-$adjointness through STSOs. The present paper focuses on this approximation in calculations on He$-$like systems. Self$-$adjoint extension for \textit{ad hoc} two$-$electron Dirac$-$Coulomb Hamiltonian is investigated through STSOs. The key benefit of using STSOs to solve the two$-$electron Dirac$-$Coulomb equation is that the LCAS method can be employed without concern about contamination from electron$-$positron continuum states during the optimization of orbital parameters \cite{Jeszenszki_2022}. The whole Dirac spectrum resulting from calculations, exhibits a clear separation between positive and negative energy continuum states. A consistent order of eigenvalues that represent the discrete energy of bound states for electrons is obtained.
\begin{table*}[!]
\caption{\label{tab:He_atom} Total DHF energies (absolute values in atomic units) for the He atom with chosen parameter $z$. Sum to $N$ terms.}
\begin{ruledtabular}
\begin{tabular}{cccccc}
$z$ & $N=1$ & $N=2$ & $N=4$ & $N=6$ & $N=8$
\\
\hline
$-0.9$ &
\textbf{2.8}47 847 140 366&
\textbf{2.86}1 534 729 790&
\textbf{2.86}1 795 530 155&
\textbf{2.861 8}12 897 510&
\textbf{2.861 813} 284 282
\\
\hline
$-0.5$ &
\textbf{2.8}47 823 441 642&
\textbf{2.86}1 535 880 360&
\textbf{2.86}1 795 326 809&
\textbf{2.861 8}12 949 664&
\textbf{2.861 813} 297 733
\\
\hline
$-0.1$ &
\textbf{2.8}47 799 746 064&
\textbf{2.86}1 538 168 085&
\textbf{2.86}1 795 216 186&
\textbf{2.861 813} 000 064&
\textbf{2.861 813 3}10 291 
\\
\hline
$\hspace{2.5mm}0$ &
\textbf{2.8}47 793 824 071&
\textbf{2.86}1 538 455 140&
\textbf{2.86}1 795 162 572&
\textbf{2.861 813} 012 701&
\textbf{2.861 813 3}21 769
\\
\hline
$\hspace{2.5mm}0.1$ &
\textbf{2.8}47 787 900 036&
\textbf{2.86}1 538 740 559&
\textbf{2.86}1 795 203 275&
\textbf{2.861 813} 027 505&
\textbf{2.861 813 3}26 373
\\
\hline
$\hspace{2.5mm}0.5$ &
\textbf{2.8}47 764 213 870&
\textbf{2.86}1 539 894 362&
\textbf{2.86}1 795 025 819&
\textbf{2.861 813} 072 868&
\textbf{2.861 813 3}33 885
\\
\hline
$\hspace{2.5mm}0.9$ &
\textbf{2.8}47 740 529 277&
\textbf{2.86}1 541 047 681&
\textbf{2.86}1 794 883 523&
\textbf{2.861 813} 121 306&
\textbf{2.861 813 34}1 075
\end{tabular}
\end{ruledtabular}
\footnotetext{2.861 813 322\cite{Tang_2013}}
\footnotetext{2.861 813 35\cite{Desclaux_1975}}
\footnotetext{\hspace{3mm}$1s(N=1)$.\\
$1s(\zeta)1s^{\prime}(\zeta^{\prime})(N=2)$. \\
$1s(\zeta)1s^{\prime}(\zeta^{\prime})2s(\zeta)2s^{\prime}(\zeta^{\prime})(N=4)$.\\
$1s(\zeta)1s^{\prime}(\zeta^{\prime})2s(\zeta)2s^{\prime}(\zeta^{\prime})3s(\zeta)3s^{\prime}(\zeta^{\prime})(N=6)$.\\
$1s(\zeta)1s^{\prime}(\zeta^{\prime})2s(\zeta)2s^{\prime}(\zeta^{\prime})3s(\zeta)3s^{\prime}(\zeta^{\prime})4s(\zeta)4s^{\prime}(\zeta^{\prime})(N=8)$.\\
The optimized values of $\left\lbrace \zeta, \zeta^{\prime} \right\rbrace$ for values of $N$, $N=1,2,4$ are used in $N=8$.
}
\end{table*}
\section{The Dirac$-$Hartree$-$Fock Equations}
\subsection{The Algebraic Approximation}
Given the aforementioned revision of formalism that should improve the accuracy, this study concentrates on investigating the self$-$adjointness of the conventional Dirac$-$Hartree$-$Fock (DHF) equations for many$-$electron atomic systems \cite{Kim_1967, Grant_2007,Malli_1975, Laaksonen_1988, Yanai_2001} while STSOs are a basis used in the algebraic approximation. \\
The wave$-$function $\Psi$ is a single anti$-$symmetrized product of atomic spinors $\psi$, 
\begin{align}\label{eq:13}
\Psi
=\frac{1}{\sqrt{N!}}\sum_{p}\left(-1 \right)^{p}P\left[
\psi_{1}\left(\vec{r}_{1} \right)
\psi_{2}\left(\vec{r}_{2} \right)
...
\psi_{N}\left(\vec{r}_{N} \right)
\right],
\end{align}
with $P$; the permutation operator. The $\psi$ are expanded by the LCAS method in terms of STSOs as,
\begin{align}\label{eq:14}
\psi_{p}
=\sum_{q}^{N} X_{q}C_{qp}.
\end{align}
Further clarification requires expanding both the large$-$ and small$-$components according to Eq. (\ref{eq:2}) as follows,
\begin{align}\label{eq:15}
\psi^{\beta \eta}_{p} =
\sum_{q}^{N} \chi^{\beta \eta}_{q}C^{\beta \eta}_{qp}.
\end{align}
The DHF self$-$consistent field equations for closed$-$shell atoms in terms of matrix elements are written in the following form:
\begin{align}\label{eq:16}
\sum_{q=1}^{N}F_{pq}C_{qi}= 
\varepsilon_{i}\sum_{q=1}^{N}S_{qp}C_{qi},
\end{align}
where, $S_{pq}$ are elements of overlap matrix, 
\begin{align}\label{eq:17}
\renewcommand*{\arraystretch}{2.0}
S_{pq}=
\begin{pmatrix} 
S_{pq}^{11} & 0 \\ 
0 & S_{pq}^{-1-1}
\end{pmatrix},
\end{align}
$C_{qi}$ are the linear combination coefficients,
\begin{align}\label{eq:18}
\renewcommand*{\arraystretch}{2.0}
C_{pq}=
\begin{pmatrix} 
C_{pq}^{11} \\ 
C_{pq}^{-1-1}
\end{pmatrix}.
\end{align}
The Fock matrix $F_{pq}$ is given by,
\begin{align}\label{eq:19}
\renewcommand*{\arraystretch}{2.0}
F_{pq}=
\begin{pmatrix} 
F_{pq}^{11} & F_{pq}^{1-1} \\ 
F_{pq}^{-11} & F_{pq}^{-1-1}
\end{pmatrix},
\end{align}
here,
\begin{align}\label{eq:20}
\renewcommand*{\arraystretch}{2.0}
\begin{matrix}
F_{pq}^{11} = V_{pq}^{11}+J_{qp}^{11}-K_{pq}^{11}
\\
F_{pq}^{1-1} = c\Pi_{pq}^{1-1}-K_{pq}^{1-1}
\\
F_{pq}^{1-1} = c\Pi_{pq}^{-11}-K_{pq}^{-11}
\\
F_{pq}^{-1-1} = V_{pq}^{-1-1}+J_{qp}^{-1-1}-K_{pq}^{-1-1}-2c^{2}S_{pq}^{-1-1}
\end{matrix} 
\end{align}
The following is a representation of the matrix elements in Eq. (\ref{eq:20}) in terms of the STSOs:\\
The overlap matrix:
\begin{align}\label{eq:21}
S_{pq}^{\beta \beta} = \sum_{\eta}
\Bigl \langle
\chi_{p}^{\beta \eta}
\Big \vert
\chi_{q}^{\beta \eta}
\Bigl \rangle
\end{align}
The nuclear attraction integral matrix:
\begin{align}\label{eq:22}
V_{pq}^{\beta \beta} = \sum_{\eta}
\Bigl \langle
\chi_{p}^{\beta \eta}
\Big \vert V\left(r_{i}\right) \Big \vert
\chi_{q}^{\beta \eta}
\Bigl \rangle
\end{align}
The non-Hermitian kinetic energy matrix is written:
\begin{align}\label{eq:23}
\Pi_{pq}^{\beta \beta^{\prime}} =
\Bigl \langle
\chi_{p}^{\beta}
\Big \vert \left( \vec{\sigma}. \hat{\vec{p}} \right) \Big \vert
\chi_{q}^{\beta^{\prime}}
\Bigl \rangle.
\end{align}
The Coulomb and exchange matrices are defined as,
\begin{align}\label{eq:24}
J_{pq}^{\beta \beta}=
\sum_{\beta^{\prime}} \sum_{rs}
R_{sr}^{\beta^{\prime} \beta^{\prime}} J_{pqrs}^{\beta \beta^{\prime}},
\end{align}
\begin{align}\label{eq:25}
K_{pq}^{\beta \beta}=
\sum_{rs}
R_{sr}^{\beta \beta^{\prime}} K_{pqrs}^{\beta \beta^{\prime}}.
\end{align}
The integrals arising in Coulomb and exchange matrices are determined by,
\begin{align}\label{eq:26}
J_{pqrs}^{\beta^{} \beta^{\prime}}
 =\sum_{\eta^{} \eta^{\prime}}
\Bigl \langle
\chi_{p}^{\beta^{} \eta^{}} \chi_{q}^{\beta^{} \eta^{}}
\Big \vert V\left(r_{ij}\right) \Big \vert
\chi_{r}^{\beta^{\prime}\eta^{\prime}}\chi_{s}^{\beta^{\prime}\eta^{\prime}}
\Bigl \rangle ,
\end{align}
\begin{align}\label{eq:27}
 K_{pqrs}^{\beta \beta^{\prime}}
 =\sum_{\eta \eta^{\prime}}
\Bigl \langle
 \chi_{p}^{\beta \eta} \chi_{s}^{\beta \eta}
 \Big \vert V\left(r_{ij}\right) \Big \vert
 \chi_{r}^{\beta^{\prime}\eta^{\prime}}\chi_{q}^{\beta^{\prime} \eta^{\prime}}
\Bigl \rangle.
\end{align}
Components of Eqs. (\ref{eq:26}, \ref{eq:27}) are the one$-$center two$-$electron integrals over non$-$relativistic, non$-$integer Slater$-$type orbitals ($n^{*}-$STOs).
\begin{table*}[htp!]
\begin{threeparttable}
\caption{\label{tab:He_like_atom} Total DHF energies (absolute value, atomic units) for He-like atom depending on parameter $z$ and upper limit of summation $N$.}
\begin{ruledtabular}
\begin{tabular}{ccccc}
$Z$ & $N=1$ & $N=2$ & $N=4$ & $N=6$ 
\\
\hline
$4$ &
\begin{tabular}{ccccc}
13.600 377 616 950\\
13.600 118 634 630
\end{tabular} &
\begin{tabular}{ccccc}
13.613 442 208 450\\
13.611 543 654 425
\end{tabular} &
\begin{tabular}{ccccc}
13.613 999 007 620\\
13.613 999 827 086
\end{tabular} &
\begin{tabular}{ccccc}
13.614 000 800 629\\
13.614 001 458 166\\
13.658 26\tnote{a}\\
13.614 001 410\tnote{c}
\end{tabular}
\\
\hline
$8$ &
\begin{tabular}{ccccc}
59.146 390 577 225\\
59.144 225 044 802
\end{tabular} &
\begin{tabular}{ccccc}
59.159 626 713 063\\
59.155 685 619 330
\end{tabular} &
\begin{tabular}{ccccc}
59.159 791 897 943\\
59.159 798 086 665
\end{tabular} &
\begin{tabular}{ccccc}
59.159 793 520 587\\
59.159 795 861 794\\
59.205 19\tnote{a}\\
59.159 794 044\tnote{c}
\end{tabular}
\\
\hline
$10$ &
\begin{tabular}{ccccc}
93.969 474 016 143\\
93.965 201 425 310
\end{tabular} &
\begin{tabular}{ccccc}
93.982 558 067 766\\
93.980 547 701 435
\end{tabular} &
\begin{tabular}{ccccc}
93.982 797 374 795\\
93.982 796 543 036
\end{tabular} &
\begin{tabular}{ccccc}
93.982 799 062 140\\
93.982 795 713 484\\
93.982 8\tnote{b}\\
93.982 799 60\tnote{d}
\end{tabular}
\\
\hline
$14$ &
\begin{tabular}{ccccc}
187.829 168 061 867\\
187.817 270 675 960
\end{tabular} &
\begin{tabular}{ccccc}
187.842 122 635 359\\
187.842 473 947 194
\end{tabular} &
\begin{tabular}{ccccc}
187.841 516 858 684\\
187.842 098 314 358
\end{tabular} &
\begin{tabular}{ccccc}
187.842 316 982 918\\
187.842 258 675 379\\
187.888 1\tnote{a}\\
187.842 328 57\tnote{c}
\end{tabular}
\\
\hline
$16$ &
\begin{tabular}{ccccc}
246.927 133 792 602\\
246.909 262 655 469
\end{tabular} &
\begin{tabular}{ccccc}
246.940 159 755 038\\
246.939 499 998 357
\end{tabular} &
\begin{tabular}{ccccc}
246.940 198 330 426\\
246.939 812 803 598\
\end{tabular} &
\begin{tabular}{ccccc}
246.940 199 974 649\\
246.939 993 999 005\\
246.940 200 3\tnote{d}
\end{tabular}
\\
\hline
$18$&
\begin{tabular}{ccccc}
314.187 200 808 282\\
314.161 597 999 158
\end{tabular} &
\begin{tabular}{ccccc}
314.199 084 622 854\\
314.200 276 768 477
\end{tabular} &
\begin{tabular}{ccccc}
314.200 162 541 732\\
314.199 320 254 577
\end{tabular} &
\begin{tabular}{ccccc}
314.200 164 181 523\\
314.199 742 082 698\\
314.246 0\tnote{a}&\\
314.200 163 58\tnote{c}
\end{tabular}
\\
\hline
$20$&
\begin{tabular}{ccccc}
389.653 916 068 479\\
389.618 580 317 670
\end{tabular} &
\begin{tabular}{ccccc}
389.665 687 123 168\\
389.667 229 037 539
\end{tabular} &
\begin{tabular}{ccccc}
389.666 761 208 922\\
389.665 129 777 430\
\end{tabular} &
\begin{tabular}{ccccc}
389.666 761 657 877\\
389.665 992 868 324\\
389.666 8\tnote{b}\\
389.666 763 2\tnote{d}
\end{tabular}
\\
\hline
$30$ &
\begin{tabular}{ccccc}
892.062 223 532 728\\
891.938 749 455 451
\end{tabular} &
\begin{tabular}{ccccc}
892.073 834 356 108\\
892.084 044 729 572
\end{tabular} &
\begin{tabular}{ccccc}
892.074 286 530 026\\
892.056 505 218 534\
\end{tabular} &
\begin{tabular}{ccccc}
892.074 288 262 968\\
892.067 664 277 909\\
892.074 3\tnote{b}\\
892.074 287 0\tnote{d}
\end{tabular}
\\
\hline
$40$ &
\begin{tabular}{ccccc}
1609.894 429 038 061\\
1609.588 220 277 417
\end{tabular} &
\begin{tabular}{ccccc}
1609.900 674 984 374\\
1609.929 700 413 896
\end{tabular} &
\begin{tabular}{ccccc}
1609.905 281 770 502\\
1609.808 525 776 256
\end{tabular} &
\begin{tabular}{ccccc}
1609.905 276 668 565\\
1609.873 694 489 409\\
1609.905 3\tnote{b}\\
1609.905 282\tnote{d}
\end{tabular}
\\
\hline
$50$ &
\begin{tabular}{ccccc}
2556.443 501 449 181\\
2555.809 274 182 230
\end{tabular} &
\begin{tabular}{ccccc}
2556.452 067 843 995\\
2556.512 170 326 102
\end{tabular} &
\begin{tabular}{ccccc}
2556.452 558 905 252\\
2556.065 170 462 154
\end{tabular} &
\begin{tabular}{ccccc}
2556.452 559 935 264\\
2556.330 827 656 454\\
2556.452 5\tnote{b}\\
2556.452 547\tnote{d}
\end{tabular}
\\
\hline
$60$ &
\begin{tabular}{ccccc}
3750.969 574 238 728\\
3749.788 342 233 996
\end{tabular} &
\begin{tabular}{ccccc}
3750.975 854 349 627\\
3751.068 184 828 664
\end{tabular} &
\begin{tabular}{ccccc}
3750.975 955 273 183\\
3749.685 270 052 264
\end{tabular} &
\begin{tabular}{ccccc}
3750.975 956 915 831\\
3750.575 030 300 709\\
3750.975 9\tnote{b}\\
3750.975 927\tnote{d}
\end{tabular}
\\
\hline
$70$ &
\begin{tabular}{ccccc}
5221.018 106 617 059\\
5218.956 199 968 598
\end{tabular} &
\begin{tabular}{ccccc}
5221.020 563 755 183\\
5221.209 499 311 129
\end{tabular} &
\begin{tabular}{ccccc}
5221.020 464 551 410\\
5217.216 512 537 061
\end{tabular} &
\begin{tabular}{ccccc}
5221.020 452 302 157\\
5219.797 492 341 645\\
5221.020 4\tnote{b}\\
5221.020 394\tnote{d}
\end{tabular}
\\
\hline
$80$&
\begin{tabular}{ccccc}
7006.450 807 761 485\\
7002.982 913 369 888
\end{tabular} &
\begin{tabular}{ccccc}
7006.449 857 623 003\\
7006.658 727 455 806
\end{tabular} &
\begin{tabular}{ccccc}
7006.446 099 330 204\\
6996.083 210 052 041
\end{tabular} &
\begin{tabular}{ccccc}
7006.446 794 674 882\\
7003.078 090 560 999\\
7006.446 7\tnote{b}\\
7006.446 748\tnote{d}
\end{tabular}
\end{tabular}
\end{ruledtabular}
    \begin{tablenotes}
            \item[a] \cite{Parpia_1990}
            \item[b] \cite{Inoue_2021}
            \item[c] \cite{Goldman_1988}
            \item[d] \cite{Tang_2013}
            \item[e]{
            $1s(N=1)$. $\zeta$ is optimized.\\
            \hspace{2mm}$1s(\zeta)1s^{\prime}(\zeta^{\prime})$$(N=4)$. $\left\lbrace \zeta, \zeta^{\prime} \right\rbrace$ are optimized.\\
            \hspace{2mm}$1s(\zeta)1s^{\prime}(\zeta^{\prime})2s(\zeta)2s^{\prime}(\zeta^{\prime})$$(N=4)$. $ \left\lbrace \zeta, \zeta^{\prime} \right\rbrace $ are optimized.\\
\hspace{2mm}$1s(\zeta)1s^{\prime}(\zeta^{\prime})2s(\zeta)2s^{\prime}(\zeta^{\prime})3s(\zeta)3s^{\prime}(\zeta^{\prime})(N=6)$. $N=4$ orbital parameters are used.
            }
        \end{tablenotes}
\end{threeparttable}
\end{table*}
\subsection{Evaluation of electrostatic integrals}
The challenge of integral evaluation using $n^{*}-$STOs has been a long$-$standing concern. The literature contains numerous research articles, some presenting questionable results. Over the past few years, the present authors have achieved notable success in addressing this issue \cite{Bagci_2014} (see also references therein). Two novel functions have been introduced, one for atomic calculations \cite{Bagci_2023} and the other for molecular calculations \cite{Bagci_2015}. These new functions are a significant improvement over previous methods and could be used to solve a wide range of problems. The $n^{*}-$STOs are given as,
\begin{align}\label{eq:28}
\chi_{n^{*}lm}\left( \vec{r}, \zeta \right)
=r^{n^{*}-1}e^{-\zeta r}S_{lm}\left(\theta, \varphi\right),
\end{align}
here, $S_{lm}\left(\theta, \varphi\right)$ are real or complex spherical harmonics $\left( S_{lm}\left(\theta, \varphi\right) \equiv Y_{lm}\left(\theta, \varphi\right) \right)$. \\
Consider the electrostatic integrals arising in Eqs. (\ref{eq:26}, \ref{eq:27}). They are expressed in terms of hyper$-$geometric functions which are difficult to compute because of non$-$trivial structure of infinite series used to define them. Through the Laplace expansion of Coulomb operator for the radial part of electrostatic interaction a relationship with the following form is derived:
\begin{multline}\label{eq:29}
R^{L}_{n^{*},n^{*\prime}}\left(\zeta,\zeta^{\prime}\right)
=\dfrac{\Gamma\left(n^{-}+n^{*\prime}+1\right)}{\left(\zeta+\zeta^{\prime}\right)^{n^{*}+n^{*\prime}+1}}
\Bigg\{
\dfrac{1}{n^{*}+L+1}
\\
\times {_2}F_{1}\left[1,n^{*}+n^{*\prime}+1,n^{*}+L+2;\frac{\zeta}{\zeta+\zeta^{\prime}}\right]
+\dfrac{1}{n^{*\prime}+L+1}
\\
\times {_2}F_{1}\left[1,n^{*}+n^{*\prime}+1,n^{*\prime}+L+2;\frac{\zeta^{\prime}}{\zeta+\zeta^{\prime}}\right]
\Bigg\}.
\end{multline}
where, $n^{*}_{1}+n^{*\prime}_{1}$ and $n^{*}_{2}+n^{*\prime}_{2}$ are replaced by $n^{*}$ and $n^{*\prime}$, then, similarly $\zeta_{1}+\zeta^{\prime}_{1}$ and $\zeta_{2}+\zeta^{\prime}_{2}$ by $\zeta$ and $\zeta^{\prime}$, respectively.\\
The hyper$-$geometric functions are expressed  in terms of power series as: \cite{Bateman_1953},\\
for $\vert z \vert < 1$,
\begin{align}\label{eq:30}
{_2}F_{1}\left[a,b,c,z\right] =\sum_{n=0}^{\infty} \frac{\left(a\right)_{n}\left(b\right)_{n}}{\left(c\right)_{n}} \frac{1}{n!} z^{n}.   
\end{align}
$\left(a\right)_{n}=\frac{\Gamma\left(a+n\right)}{\Gamma\left(a\right)}$ is the Pochhammer symbol. $\Gamma\left(a\right)$ is the gamma function. Very numerous terms arising from the electrostatic interaction within the many$-$electron Dirac$-$Fock equation poses a formidable obstacle when attempting to employ either power series representations or recurrence relationships for hyper$-$geometric functions. An alternative and more efficient formula has been derived for Eq. (\ref{eq:29}). New Hyper$-$radial functions based on the bi$-$directional method have been introduced. Through the bi$-$directional method the hyper$-$geometric functions are avoided from electrostatic integrals as \cite{Bagci_2023},
\begin{multline}\label{eq:31}
{_2}F_{1}\left[1,n+n^{\prime}+1,n+L+2;\frac{\zeta}{\zeta+\zeta^{\prime}}\right]
\\
=\dfrac{R^{L}_{n,n^{\prime}}\left(\zeta, \zeta^{\prime}\right)+m^{L}_{n^{\prime}n}\left(\zeta^{\prime},\zeta\right)}{e^{L}_{nn^{\prime}}h_{n^{\prime}n}^{L}\left(\zeta^{\prime},\zeta\right)},
\end{multline}
\begin{multline}\label{eq:32}
{_2}F_{1}\left[1,n+n^{\prime}+1,n^{\prime}+L+2;\frac{\zeta^{\prime}}{\zeta+\zeta^{\prime}}\right]
\\
=\dfrac{R^{L}_{n^{\prime},n}\left(\zeta^{\prime}, \zeta\right)+m^{L}_{nn^{\prime}}\left(\zeta, \zeta^{\prime}\right)}{e^{L}_{n^{\prime}n}h_{nn^{\prime}}^{L}\left(\zeta, \zeta^{\prime}\right)},
\end{multline}
In following they are transformed to hyper$-$radial functions,
\begin{align}\label{eq:33}
\mathfrak{R}^{L}_{n,n^{\prime}}\left(\zeta,\zeta^{\prime}\right)
=\dfrac{R^{L}_{n,n^{\prime}}\left(\zeta,\zeta^{\prime}\right)+m^{L}_{n^{\prime}n}\left(\zeta^{\prime}, \zeta\right)}{e^{L}_{nn^{\prime}}h_{n^{\prime}n}^{L}\left(\zeta^{\prime},\zeta\right)}.
\end{align}
These functions can simply be calculated as,
\begin{multline}\label{eq:34}
\mathfrak{R}^{L+2}_{n,n^{\prime}}\left(\zeta,\zeta^{\prime}\right)
=\dfrac{\left(n+L+3\right)}{\zeta\left(n+L+2\right)\left(-n^{\prime}+L+2\right)}
\\
\times \bigg\{
\zeta^{\prime} \left(n+L+2\right)\mathfrak{R}^{L}_{n,n^{\prime}}\left(\zeta,\zeta^{\prime}\right)
\\
+\left[\zeta \left(-n^{\prime}+L+1\right)-\zeta^{\prime}\left(n+L+2\right)\right]
\\
\times
\mathfrak{R}^{L+1}_{n,n^{\prime}}\left(\zeta,\zeta^{\prime}\right)
\bigg\},
\end{multline}
where,
\begin{multline}\label{eq:35}
\mathfrak{R}^{0}_{n,n^{\prime}}\left(\zeta,\zeta^{\prime}\right)=
\dfrac{R^{0}_{n,n^{\prime}}\left(\zeta,\zeta^{\prime}\right)+m^{0}_{n^{\prime}n}\left(\zeta^{\prime},\zeta\right)}{e^{0}_{nn^{\prime}}h_{n^{\prime}n}^{0}\left(\zeta^{\prime}, \zeta\right)}
\\
=\left(n+1\right)\left(\dfrac{\zeta}{\zeta+\zeta^{\prime}}\right)^{-n-1}
\left(\dfrac{\zeta^{\prime}}{\zeta+\zeta^{\prime}}\right)^{-n^{\prime}}
\\
\times \Bigg\{
\dfrac{\Gamma\left(n+1\right)\Gamma\left(n^{\prime}\right)}{\Gamma\left(n+n^{\prime}+1\right)}
-B_{n^{\prime}n+1}\left(\dfrac{\zeta^{\prime}}{\zeta+\zeta^{\prime}}\right)
\Bigg\},
\end{multline}
and,
\begin{multline}\label{eq:36}
\mathfrak{R}^{1}_{n,n^{\prime}}\left(\zeta,\zeta^{\prime}\right)=
\\
\left(\dfrac{n+2}{n^{\prime}-1}\right)
\bigg[
\left(\dfrac{\zeta^{\prime}}{\zeta}\right)
\mathfrak{R}^{0}_{n^{\prime},n}\left(\zeta,\zeta^{\prime}\right)
-\left(\dfrac{\zeta+\zeta^{\prime}}{\zeta}\right)
\bigg].
\end{multline}
with $B_{nn^{\prime}}$ are the incomplete beta functions.
\section{Results and Discussions}
A self$-$consistent field procedure for Eq.(\ref{eq:16}) is implemented via a computer program code written in the Mathematica programming language \cite{Mathematica_2021}. The challenges that should be noted to obtain these first test results of using STSOs in calculation of many$-$electron systems are related to  programming, rather then methodology. The computer program constructed in Mathematica for DHF calculations is based on functions and modules. The non$-$relativistic one$-$ and two$-$electron integrals are defined as functions. Modules are used for their relativistic counterparts. In addition, several modules are integrated for optimization and $SCF$ procedures. The resulting matrices are too large, thus the command $\$MaxExtraPrecision$ has been required in numerous sections of the program. These creates a disadvantage both in terms of computational speed and precision of numerical values. To ensure a computation time within acceptable limits, the upper limit of summation in LCAS is chosen as $N=8$.\\
Variationally optimum values for $\zeta$ are obtained by the Rayleigh$-$Ritz variational method. The same basis set approximation as in \cite{Datta_1994} is employed. The values for principal quantum numbers are determined by,
\begin{align}
n^{*}=\sqrt{\kappa^2-\left(\alpha Z\right)^{2}\left(2z-1\right)}
\end{align}
where, the parameter $z$ posses a range of values for which the principal quantum number $n^{*}$ is positive. Variationally minimum values for the total energies of two$-$electron atoms are presented in Tables \ref{tab:He_atom} and \ref{tab:He_like_atom}. The results are presented for nuclear charges ranging from $Z=2$ to $Z=80$, depending on the values of $z$. In these tables, we present a comparison with results reported in \cite{Parpia_1990, Inoue_2021, Goldman_1988, Tang_2013}, which in most cases were calculated using the \textit{General Relativistic Atomic Structure Program} (GRASP) \cite{Dyall_1989}. It can be seen from these tables that no situation that triggers Brown$-$Revenhall disease has been identified. Nevertheless, following the no$-$pair approach, as the nuclear charge increases, it becomes evident that the precision derived from the calculation results diminishes, occasionally yielding outcomes inferior to those attained through numerical methods or $CI$ expansion. This conclusion has also been reported in \cite{Tang_2013}. Fixed values for parameter $z$ are used in the present study. In Table \ref{tab:He_like_atom}, the first column lists values of $z$, $z=0$, and the second column lists values of $z$, $z=0.5$. The convergence property of this table reflects the graphics presented in Figure 2 of \cite{ABagci_2020}, where the results oscillate for non-zero $z$ as $N$ increases, but eventually converge to the exact result. The tables show that $z$ can be assigned as a variational parameter. This may lead to more accurate results for the total energies of atoms with high nuclear charge. For any value of $z$ as the upper limit of summation increases the convergence is satisfied (Table \ref{tab:He_atom}).\\
L{\"o}wdin's symmetric orthogonalization method \cite{Lowdin_1950} is used in this study. An alternative solution to the generalized eigenvalue problem must be sought. The Mathematica calculations become impractical for high values of the upper limit of the summation $N$. Such a study, using $z$ as a variational parameter, is planned for the future.

\end{document}